\documentclass[
 reprint,amsmath,amssymb,aps,
prb,
floatfix,
]{revtex4-2}%\documentclass[aps,prb,twocolumn,12pt]{article}
\usepackage{xcolor}  % In the preamble
\usepackage{placeins}
\usepackage{soul}
\usepackage{graphicx,float}
\usepackage{amsmath,bm}
\usepackage{amssymb}
\usepackage{braket}
\usepackage{amsfonts}
\usepackage{dsfont}
\usepackage{comment,color}
\usepackage{lipsum}
\usepackage{hyperref}
\usepackage{textcomp}
\hypersetup{colorlinks=true,linkcolor=blue,anchorcolor=blue,citecolor=blue,filecolor=blue,urlcolor=blue,bookmarksnumbered=true,pdfview=FitB}
\begin{document}
%\doublespacing

\title{Intrinsic magnetotransport and orientation dependent topological Hall effect in EuAuBi}
\author{Lipika$^{1}$}
\author{Sneh$^{1}$}
\author{Shobha Singh$^1$}
\author{Ralf Koban$^2$}
\author{Walter Schnelle$^2$}
\author{Kaustuv Manna$^{1,*}$}
\affiliation{$^1$Department of Physics, Indian Institute of Technology Delhi, Hauz Khas, New Delhi, India 110016}
\affiliation{$^2$ Max-Planck-Institute for Chemical Physics of Solids,Nöthnitzer Straße 40
01187 Dresden, Germany}
\affiliation{$^*$kaustuvmanna@physics.iitd.ac.in}
\

\date{\today}
      
\begin{abstract}
We report the growth of high-quality single crystals of the magnetic topological semimetal EuAuBi using a Pb-flux method, which effectively suppresses the formation of secondary Au$_2$Bi impurity phases that were prevalent in the previously reported Bi-flux grown crystals. This growth optimization enables reliable investigation of the intrinsic physical properties of EuAuBi. Importantly, Pb flux growth stablilizes \textit{c}-axis–oriented single crystals, enabling Hall measurements in a previously unexplored geometry. In this configuration  (\textit{I} $\parallel$ \textit{c}, \textit{B} $\perp$ \textit{c}), a finite residual Hall contribution, known as topological Hall signal emerges below the antiferromagnetic ordering temperature and within a narrow magnetic field-range. This Hall contribution coincides with the metamagnetic transitions, anomalies in magnetoresistance, and an additional feature in field-dependent specific heat, indicating strong coupling between the electronic transport and field-induced magnetic reconstructions. Consequently, these findings underscore the significance of crystal orientation in uncovering topological transport signatures in magnetic semimetals such as EuAuBi. 

\end{abstract}

\maketitle  
 Topological semimetals are an important class of quantum materials, characterized by nontrivial band crossings near the Fermi level, which results in a finite Berry curvature. This acts as a fictitious magnetic field in momentum space, leading to various exotic phenomena, including the anomalous Hall effect, anomalous Nernst effect, chiral anomaly, etc \cite{Armitage2018, Lv2021, Yan2017, Manna2018, Wang2017}.
Magnetic topological materials are of particular interest as magnetic order introduces an additional degree of freedom for Berry curvature engineering \cite{Tokura2019, Bernevig2022, Nakatsuji2022}. Consequently, these materials hold significant potential for applications in spintronics, energy conversion, and quantum devices \cite{Duan2024, Yang2025, Luo2022, Li2021TopologicalStorage}.
\newline
Beyond momentum-space topology, the non-collinear spin textures such as skymrions can induce a non-zero real-space Berry curvature, which acts as an emergent magnetic field proportional to the spin chirality \textit{S$_1$(S$_2$$\times$S$_3$}). This field interacts with the conduction electrons, resulting in topological Hall effect \cite{SKYRME1962556,Nagaosa2013,Srgers2014,Neubauer2009TopologicalMnSi}. While such magnetic textures are commonly stabilized in non-centrosymmetric systems via the Dzyaloshinskii-Moriya interaction (DMI) \cite{Neubauer2009TopologicalMnSi,doi:10.1126/science.1166767,Zhang2016AntiferromagneticManipulation,doi:10.1126/science.1214143}, recent studies indicate that centrosymmetric lattices can also support these non-trivial magnetic textures through mechanisms such as competing exchange interactions, magnetic frustration, or anisotropy \cite{Li2019LargeFe3Sn2, Ghimire2020}. However, material systems that simultaneously host nontrivial topology in both real-space and momentum-space remain rare.
\newline
 Recent studies have identified EuAuBi as an antiferromagnetic topological semimetal posited to exhibit non-trivial topology in both momentum and real space \cite{Lipika2026}. Theoretical density functional theory (DFT) calculations suggest that EuAuBi is an easily tunable topological semimetal, capable of transitioning to a triple-degenerate nodal point semimetal under an applied magnetic field \cite{Chi2024ElectronicBi}. Previous experimental studies have documented a complex magnetic phase diagram of this system, with features in ac susceptibility suggesting the possible emergence of field-induced nontrivial real-space magnetic textures, although no definitive signature of topological Hall effect has been observed \cite{Takahashi2023SuperconductivityEuAuBi,Lipika2026}. EuAuBi has also been reported to exhibit unconventional superconductivity with \textit{T$_c$} $\sim$ 2 K \cite{Takahashi2023SuperconductivityEuAuBi}. Additionally, first-principles calculations have highlighted its potential as a promising material for ferroelectric applications \cite{Tan2025PromisingCurrent}.
 \newline
 \begin{figure}[t]
{  
    \includegraphics[width=8.5 cm]{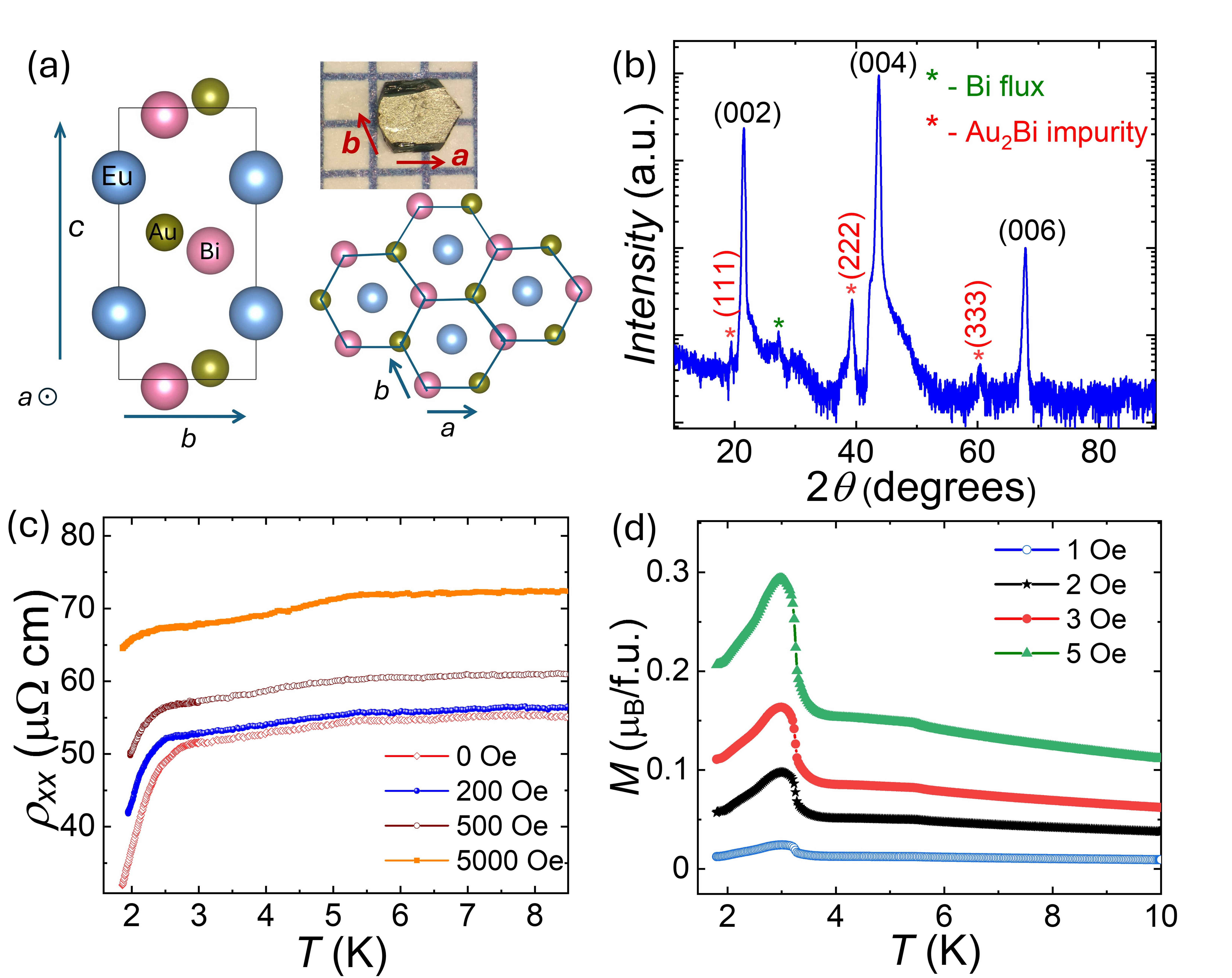 }
    \caption{(a) Crystal structure of EuAuBi, the blue, pink, and brown represents Eu,Bi, and Au atoms respectively. The hexagonal shaped crystal (grown using Bi flux) of approximately 2mm is shown (b) X-ray diffraction $\theta$ - 2$\theta$ plot 
    (in logarithmic scale) of EuAuBi crystal grown using Bi flux, the primary peaks corresponds to the (001) planes of EuAuBi and the peaks indicated by red asterisk corresponds to the Au$_2$Bi impurity phase formed in the crystal, Temperature-dependent (c) resistivity, and  (d) magnetization in the presence of different magnetic fields. }
    \label{fig:my_label5}
    }
\end{figure}
 In this study, we report a Pb-flux optimized growth process to produce high-quality EuAuBi single crystals to uncover its intrinsic features. We demonstrate that the previously reported superconductivity in EuAuBi is of extrinsic origin, attributable to secondary impurity
phases rather than an intrinsic bulk superconducting
state. In addition to suppressing the impurity phase, the optimization of EuAuBi crystal growth using Pb flux facilitates \textit{c}-axis oriented crystal growth, enabling the Hall measurement along a previously unexplored orientation. In this configuration, an additional contribution to the Hall effect, consistent with the topological Hall effect is observed. The absence of this contribution in the earlier reports, suggests the orientation-dependent presence of the nontrivial real-space magnetic textures in EuAuBi.
\newline
Single crystals reported in this work were grown using the flux method, employing two distinct fluxes - Bi and Pb. This research investigated various growth conditions to enhance the grown crystal quality and suppress the formation of the intermetallic binary phase Au$_2$Bi as detailed in the supplementary material. High-purity elements, Eu (Alfa Aesar, 99.99 \%), Au (99.99\%), Pb (Alfa Aeser, 99.99\%), and Bi (99.99\%) were combined in predetermined ratios within an argon-filled glove box, and subsequently vacuum sealed in a quartz tube. The ampoules were placed vertically in a muffle furnace subjected to specific temperature profiles. The X-ray diffraction experiment was performed at room temperature using HR-XRD (PANalytical Empyrean X-Ray diffractometer) with a Cu K$\alpha$ (\textit{$\lambda$} = 1.5406 \AA). The elemental composition of the crystal was verified using the Energy dispersive X-ray spectroscopy (EDX) on Hitachi- High Technol TM 3000 operating with 15 keV energy. Laue x-rays diffraction experiment was performed using the Proto HT Single crystal orientation system. Specific heat and electrical transport measurements were conducted using the Quantum Design made PPMS system, while magnetic measurements were performed using a Quantum Design made SQUID (MPMS3) magnetometer. 
\newline
\begin{figure*}[t]
{ 
  \centering  
    \includegraphics[width=17cm]{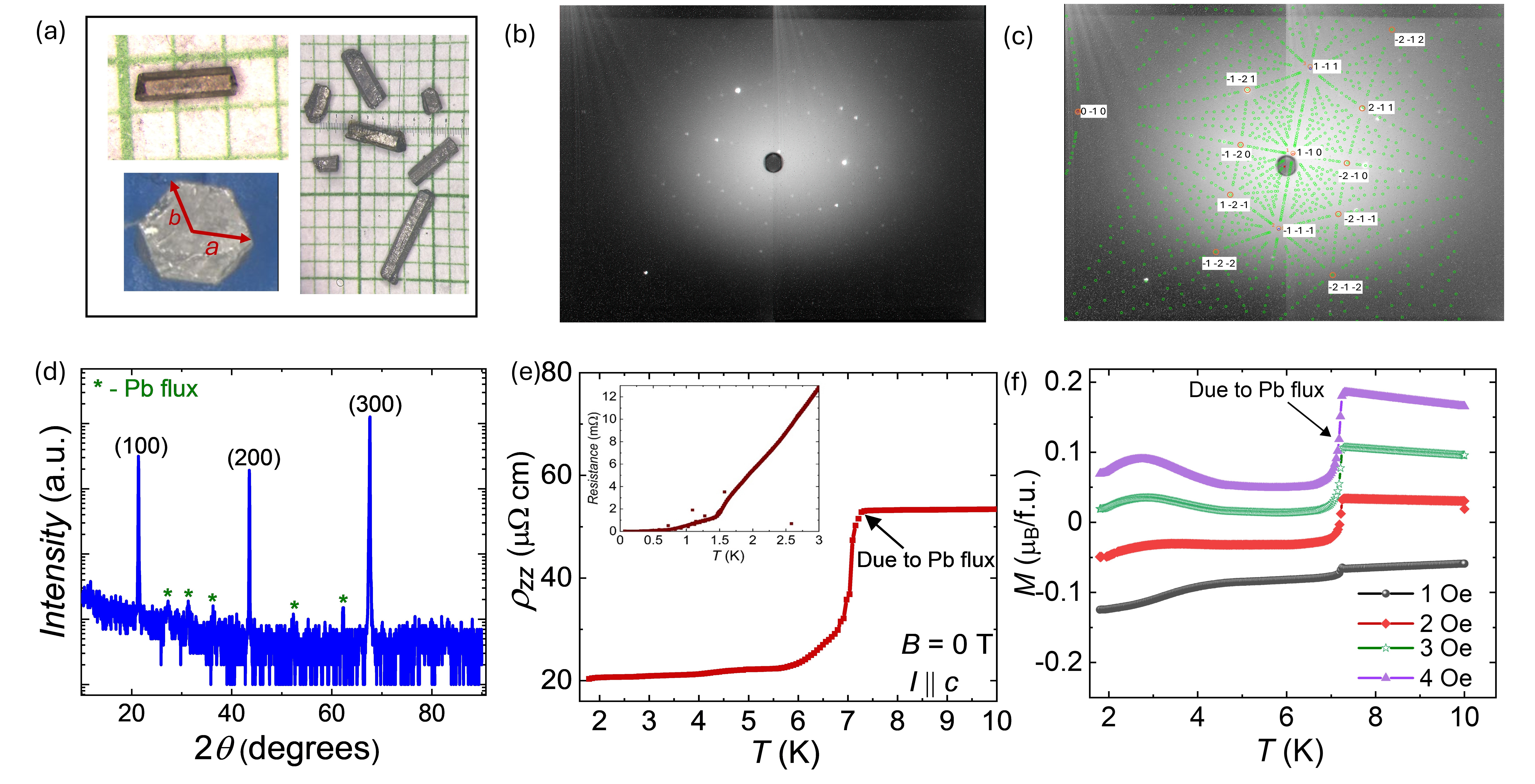 }
    \caption{(a) The as-grown rod-shaped crystals (grown using Pb flux) of approximately 2-3 mm are obtained with hexagonal cross-section area, (b) the Laue x-rays diffraction pattern of the crystal, (c) simulated pattern of the laue diffraction matches with the (1-10) plane, (d) the $\theta$ - 2$\theta$ XRD diffraction pattern shows the (\textit{h}00) plane of the crystal indicating the crystal is \textit{c}-oriented, (e) Temperature-dependent resistivity of EuAuBi crystal grown using Pb flux, the drop at 7 K is due to Pb flux \cite{1091958PHLead} that is present over the crystal surface. The inset shows the temperature-dependent resistance of the same crystal measured up to 300 mK. (f) The temperature-dependent magnetization in the presence of different fields shows the drop due to Pb flux. }
    \label{fig:my_label5}
    }
\end{figure*} 
EuAuBi crystallizes in a hexagonal crystal structure, P6$_3$\textit{mc} (space group: 186), as shown in Fig. 1(a). The atomic arrangements along the \textit{bc} and \textit{ab} planes are illustrated. The Au and Bi atoms form the  honeycomb-like structure along the \textit{ab} plane, while the Eu atoms are situated between these planes along the crystallographic \textit{c} axis. 
In this study, we initially present the physical properties of EuAuBi single crystals grown using Bi flux [Fig. 1(a) inset] and subsequently compare them with samples grown using Pb flux [Fig. 2(a) inset]. Fig. 1b shows the X-ray diffraction (XRD) pattern (intensity plotted on a logarithmic scale) of a EuAuBi single crystal grown using the Bi flux. The diffraction pattern shows the (00\textit{l}) peaks of the crystal, indicating that the crystal is \textit{ab}-oriented and the \textit{c}-axis is perpendicular to the surface. In addition to the EuAuBi reflections, impurity peaks corresponding to the (111) plane of Au$_2$Bi are also observed, confirming the presence of a secondary phase. Despite extensive optimization attempts, it remains challenging to completely eliminate the Au$_2$Bi impurity phase in the Bi-flux grown EuAuBi crystals, as detailed in the S1.1 section of the supplementary material.
Figs. 1(c) and 1(d) present the temperature-dependent electrical resistivity and magnetization measurements of the same crystal. A drop in resistivity is observed in $\rho$$_x$$_x$ around 2 K, which has been previously attributed to superconductivity in EuAuBi \cite{Takahashi2023SuperconductivityEuAuBi}. However, the corresponding magnetization data does not exhibit any features indicative of Meissner effect, even under very low applied magnetic fields, leading to suspicious intrinsic bulk superconductivity in the system. Notably, Au$_2$Bi has been reported to be a superconductor with a critical temperature \textit{T}$_c$ $\sim$ 2 K; Meissner effect in grown Au$_2$Bi crystals is also shown in the supplementary material \cite{DSHOENBERG1938, SupplementalMaterial}. This suggests that the observed superconductivity might be due to Au$_2$Bi impurity phase rather than an intrinsic property of EuAuBi.
\newline
\begin{figure*}[t]
{
    \includegraphics[width=17cm]{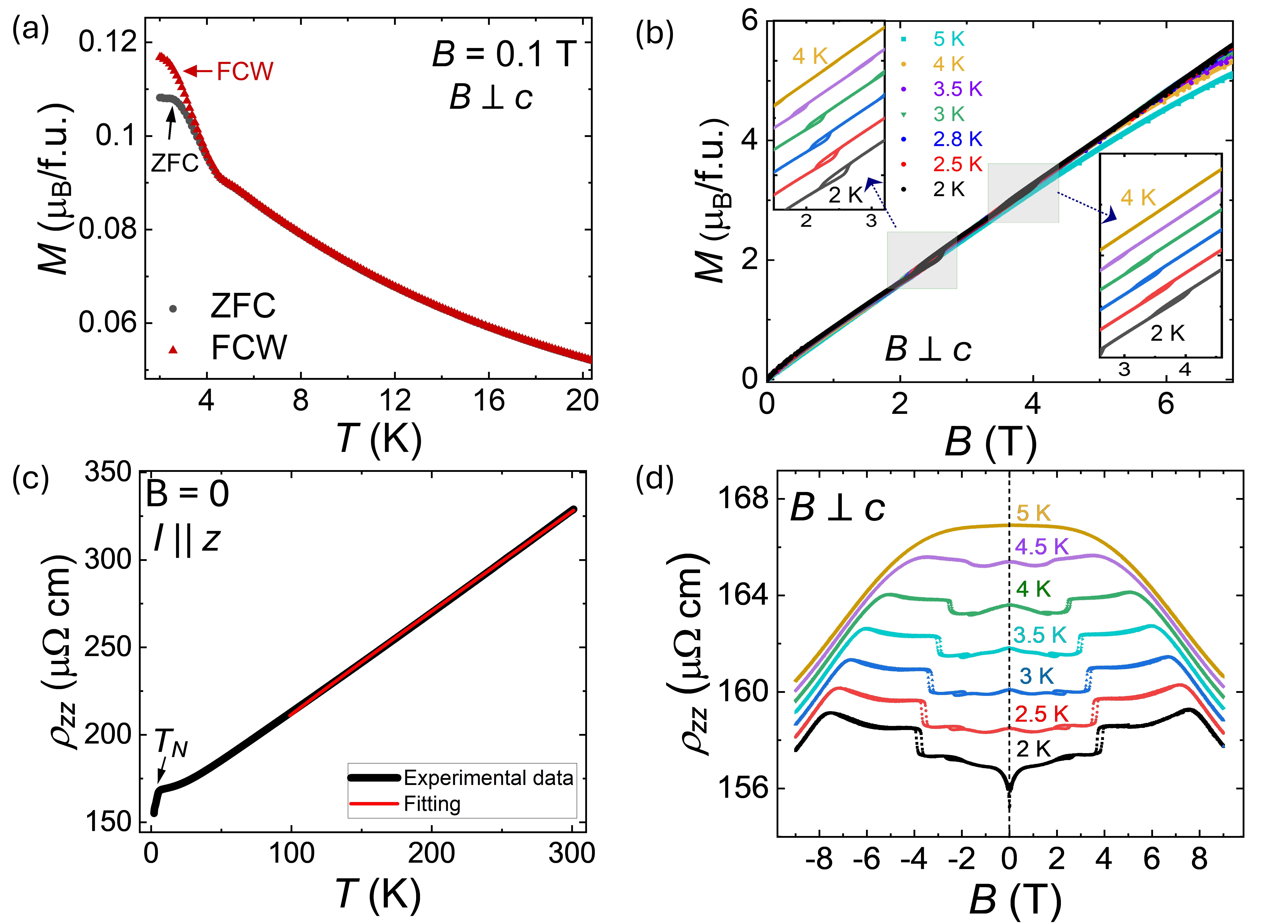}
    \caption{(a) Zero-field cooling (ZFC) and Field-cooled warming (FCW) magnetization curves in the presence of 0.1 T magnetic field applied in-plane of the crystal. (b) Field-dependent magnetization \textit{M}(\textit{B}) at different temperatures for \textit{B} $\perp$ \textit{c}. The zoomed in parts of the anomalies are shown in the insets. (c) Temperature-dependent resistivity in the absence of magnetic field for \textit{I} $\parallel$ \textit{z} is depicted in black color along with the fitting (by red color) obtained using the Eq. 1 in temperature range 100 - 300 K. (d) Field-dependent resistivity \textit{$\rho_{zz}$}  at different temperatures for \textit{B} $\perp$ \textit{c}.}
    \label{fig:my_label5}
    }
\end{figure*}
Motivated by this, we utilized Pb flux growth to suppress the formation of Au$_2$Bi, thereby enabling the investigation of intrinsic properties of EuAuBi.
Interestingly, the crystals obtained through Pb flux growth are rod-shaped, \textit{c}-axis oriented single crystals, typically measuring 2-3 mm in length, in contrast to the \textit{ab} plane-oriented plate-like crystals produced using Bi flux. The as-grown Pb flux crystals are depicted in Fig. 2(a). The crystals are elongated along the \textit{c} direction with a hexagonal cross-section, demonstrating that the choice of flux plays an important role in determining the growth orientation. The elemental composition and distribution mapping (determined by EDX) of the same batch of the EuAuBi crystal is presented in the section S2 of the supplementary material \cite{SupplementalMaterial}. Fig. 2b presents the Laue diffraction pattern of the Pb-flux grown crystal, where the well-defined Laue diffraction spots indicate the high-crystalline quality of the sample without any domains or twinning. The indexed Laue pattern, shown in Fig. 2(c), corresponds to (1-1 0) plane, as the x-ray in the Laue geometry of the system is incident to the side plane of the \textit{c}-axis oriented crystal. X-ray diffraction (Fig. 2(d)) further confirms the orientation, displaying reflections corresponding to (\textit{h}00) planes. Importantly, impurity peaks associated with  Au$_2$Bi are not observed, indicating that optimized crystal growth using Pb-flux effectively suppresses the secondary phase of Au$_2$Bi  (see supplementary material \cite{SupplementalMaterial}, S1 section for detailed optimization workflow). 
Figs. 2(e) and 2(f) present the temperature-dependent electrical resistivity and magnetization of the sample. Importantly, no drop in resistivity is observed near 2 K, and no superconducting transition is detected even at temperature as low as 300 mK, as depicted in the inset. A decrease in resistivity is observed near 7 K, which is attributed to the residual Pb flux present on the crystal surface (corroborated by the XRD pattern) \cite{1091958PHLead}. However, no superconducting transition associated with the EuAuBi crystal itself is detected. Consistently, \textit{M}(\textit{T}) exhibits a weak superconducting signal originating from residual Pb flux, but no Meissner effect corresponding to bulk EuAuBi around 2 K is observed. These findings confirm that EuAuBi does not exhibit intrinsic superconductivity. Therefore, in the rest of this study, we focus on the magnetic and transport properties of Pb-grown \textit{c}-oriented EuAuBi single crystals.
\newline 
To understand the magnetic properties of the system, we performed the temperature- and field-dependent magnetization measurements. Fig. 3(a) presents the zero-field cooled (ZFC) and field-cooled warming (FCW) magnetization data, measured under an applied field of 1000 Oe. The crystal exhibits an antiferromagnetic transition at a Neel temperature \textit{T$
_N$} $\sim$ 4.8 K, in contrast to 4 K previously reported in Bi-flux grown sample \cite{Lipika2026,Takahashi2023SuperconductivityEuAuBi}. The observed increase in the magnetic transition temperature is likely
attributable to the improved crystalline quality of the Pb-flux grown crystals, which results in reduced disorder and
defect concentration, as observed in several other material systems \cite{Attanayake2023,Li2017, Patel2022}.  The bifurcation in ZFC and FCW indicates the presence of competing interactions within the system.
Fig. 3(b)  illustrates the isothermal magnetization \textit{M}(\textit{B}) of the EuAuBi crystal measured at various temperatures with the magnetic field oriented within the \textit{ab} plane of the crystal (\textit{M}(\textit{B}) for \textit{B} $\parallel$ \textit{c} is presented in the S3 section of Supplementary material \cite{SupplementalMaterial}). The \textit{M}(\textit{B}) curves display two distinct hystereses, which shift towards lower field with increasing temperature. This behaviour is qualitatively similar to that previously reported for Bi-flux grown EuAuBi crystals, indicating that the fundamental magnetic interactions are preserved irrespective of the growth with different flux \cite{Takahashi2023SuperconductivityEuAuBi,Lipika2026}. However, the field range over which the metamagnetic transitions occur is slightly shifted compared to Bi-flux-grown crystals. This variation may be due to the demagnetization factor resulting from the different crystal morphology and orientation of the Pb grown rod-shaped crystals.
\newline
To elucidate the influence of magnetic spin configurations on the charge transport, we performed temperature- and field-dependent resistivity measurements.
Fig. 3(c) shows the temperature-dependent resistivity measured in a zero external magnetic field, with electric current applied along the \textit{c}-axis of the crystal. The resistivity decreases monotonically with decreasing temperature, indicating metallic behaviour. To determine the dominant scattering in high-temperature region (100 K - 300 K), the resistivity data was fitted using 
\begin{equation}
\rho(T) = \rho_{0} + \rho_{\mathrm{\textit{BG}}}(T)
\end{equation}
 where $\rho_{0}$ represents the temperature-independent residual resistivity arising from impurity and defect scattering,  $\rho_{\textit{BG}}$(\textit{T}) denotes the resistivity contribution due to electron-phonon interactions using the Bloch--Gr\"uneisen (BG) prediction. The resistivity contribution due to the spin-order scattering is neglected in this temperature range, as magnetic correlations should be absent well above the antiferromagnetic transition temperature. 
$\rho_{BG}$(\textit{T}) is given by :
\begin{equation}
\rho_{\mathrm{\textit{BG}}}(T) = A \left( \frac{T}{\Theta_D} \right)^5 
\int_{0}^{\Theta_D/T} 
\frac{x^5}{(e^{x}-1)(1-e^{-x})}\, dx
\end{equation}
where A is the electron-phonon coupling prefactor, and \textit{${\Theta_D}$} is the Debye temperature. The fitting of Eq.(1) to the resistivity data in the temperature range 100 - 300 K yields
$A = (3.04 \pm 0.09)\times 10^{-4}\ \Omega\,\mathrm{cm}$, 
$\rho_{0} = (1.61 \pm 0.01)\times 10^{-4}\ \Omega\,\mathrm{cm}$, and 
$\Theta_{D} = (134.96 \pm 3.62)\ \mathrm{K}$.
\begin{figure*}
{
    \centering   
    \includegraphics[width=17cm]{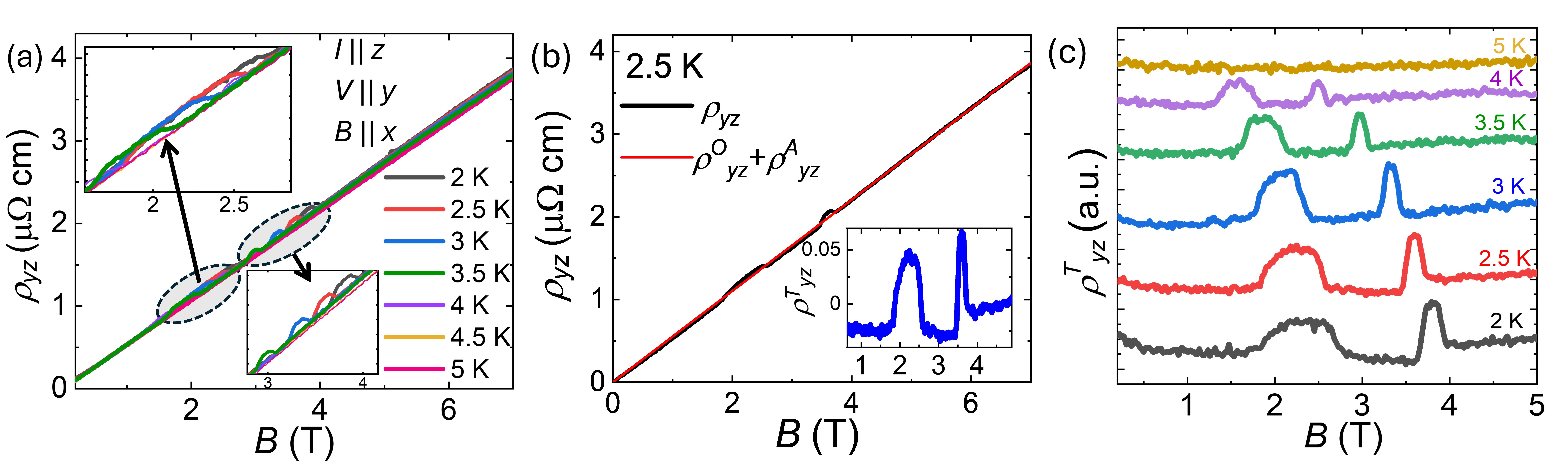}
    \caption{(a) Field-dependent Hall signal \textit{$\rho_{yz}$} at different temperatures for \textit{I} $\parallel$ \textit{z}, \textit{V} $\parallel$ \textit{y}, and \textit{B} $\parallel$ \textit{x}, (b) Fitting of \textit{$\rho_{yz}$} using \textit{$\rho$$^O_{yz}$} + \textit{$\rho$$^A_{yz}$} at \textit{T} = 2.5 K. Inset shows the topological Hall effect signal extracted at \textit{T} = 2.5 K (c) Topological Hall signal, \textit{$\rho$$^T_{yz}$} extracted from the \textit{$\rho_{yz}$} by subtracting \textit{$\rho_{yz}$}$^O$ and \textit{$\rho_{yz}$}$^A$ at different temperatures.}
    \label{fig:my_label5}
    }
\end{figure*}
\newline
In the low temperature regime, the kink-like feature is observed around 5 K [Fig.3(c)], which signifies enhanced carrier scattering arising from critical spin fluctuations in the vicinity of the antiferromagnetic transition temperature ($T_N$). Below T$_N$, the resistivity decreases rapidly, reflecting the suppression of spin-disorder scattering as the system achieves long-range magnetic order. The residual resistivity ratio (RRR), defined as $\rho$(300 K)/$\rho$(2 K), is estimated to be 2.14, indicative of good crystalline quality. This RRR value is comparable to those reported for related Eu-based ternary compounds,such as EuCuBi (RRR = 1.79) and EuCuAs (RRR = 3.3) \cite{Wang2023StructureEuCuBi, Roychowdhury2023InterplayEuCuAs}.
The field-dependent resistivity measurement [$\rho$$_{zz}$(\textit{B})] is performed at different temperatures for \textit{B} $\perp$ \textit{c} as depicted in Fig. 3(d). The observed step-like changes in the resistivity at $\sim$ 2 T and 3 T accompanied by hysteresis indicate irreversibility between the upward and downward field sweeps. These anomalies coincide with the metamagnetic transition, indicating a strong coupling between the charge transport and magnetic spin configuration. Such hysteresis in the resistivity suggests possible existence of metastable magnetic states. At higher temperatures, these features shift and weaken until they disappear above the antiferromagnetic transition temperature, confirming their magnetic origin.  The observed high-field negative magneto-resistance is likely due to reduced spin-dependent carrier scattering resulting from the polarization of the magnetic spins.
\newline
Earlier research on EuAuBi single crystals synthesized using Bi flux revealed the important features in \textit{ac} susceptibility, particularly a glassy frequency-dependent peak shift in a particular temperature and field windows with a characteristic time scale $\sim$ 10$^{-7}$s and field-induced first order phase transition in the \textit{dc} magnetization, suggesting the possible presence of nanoscale non-collinear or non-trivial magnetic textures within the system \cite{Lipika2026}. However, the absence of a topological Hall signal in the measured configuration raises question regarding the proposed magnetic textures \cite{Takahashi2023SuperconductivityEuAuBi, Lipika2026}. In this regard, the rod-shaped geometry of the Pb-flux grown EuAuBi crystals, with the length along [001] direction, enabled Hall measurements in an alternative geometry. In this setup, the current is applied along the \textit{z} direction (\textit{I} $\parallel$ z), while the applied magnetic field and Hall voltage are measured in-plane \textit{V$_H$} $\parallel$ \textit{y} and \textit{B} $\parallel$ \textit{x}. The Hall resistivity ($\rho$$_{yz}$) at various temperatures is obtained via antisymmetrization, (\textit{$\rho_{xy}$}(+\textit{B}) - \textit{$\rho_{xy}$}(-\textit{B}))/2, to remove any MR contribution due to minor misalignment of voltage probes.
 Notably, the behaviour of $\rho$$_{yz}$ shown in Fig. 4(a) is distinctly different from that of $\rho$$_{zx}$ as reported in \cite{Takahashi2023SuperconductivityEuAuBi, Lipika2026}.
\newline
The total Hall resistivity signal, \textit{$\rho_{yz}$}, can be expressed as : \textit{$\rho_{yz}$} = \textit{$\rho^O_{yz}$} + \textit{$\rho^A_{yz}$} + \textit{$\rho^T_{yz}$}, where \textit{$\rho^O_{yz}$}, \textit{$\rho^A_{yz}$}, and \textit{$\rho^T_{yz}$} correspond to the ordinary Hall effect (OHE), anomalous Hall effect (AHE), and topological Hall effect (THE), respectively. The OHE term is linear in field, given by \textit{$\rho^O_{yz}$} = \textit{R$_0$B}, where \textit{R$_0$} is ordinary Hall coefficient and AHE component is empirically modelled as : \textit{$\rho^A_{yz}$} = \textit{S}$_1$\textit{$\rho_{zz}$}\textit{M} + \textit{S}$_2$\textit{$\rho_{zz}$}\textit{$^2$M}, where  \textit{S}$_1$ and  \textit{S}$_2$ are field independent parameters, \textit{M} is isothermal magnetization, and \textit{$\rho_{zz}$} is longitudinal resistivity. The first term (\textit{S}$_1$\textit{$\rho_{zz}$}\textit{M}) corresponds for the extrinsic skew scattering mechanism, while the second term (\textit{S}$_2$\textit{$\rho_{zz}$}\textit{$^2$M}) accounts for the intrinsic and side jump scattering mechanism.
Given that, the topological Hall effect should diminish at high magnetic fields, the combined \textit{$\rho^O_{yz}$} and \textit{$\rho^A_{yz}$} contributions were  obtained by fitting the high-field Hall data (above 5 T). The best fit of $\rho_{yz}$ of high-field was achieved using the OHE and the \textit{S}$_2$\textit{$\rho_{zz}$}\textit{$^2$M}, that captures the field-dependent changes in magnetoresistance and magnetization. The alternate fittings using a constant magnetization coefficient and \textit{S}$_1$\textit{$\rho_{zz}$}\textit{M} term were also examined and are presented in the S4 section of the supplementary material \cite{SupplementalMaterial}. Using these fit parameters, the full-range Hall resistivity data was fitted at 2.5 K as illustrated in Fig. 4(b). Nevertheless, a finite Hall signal remains that cannot be explained by the OHE and AHE terms. The residual Hall signal, was extracted by subtracting the  \textit{$\rho^O_{yz}$} + \textit{$\rho^A_{yz}$} from \textit{$\rho_{yz}$} as shown in Fig. 4(b) inset. 
This signal is consistent with the topological Hall effect (THE) that persists up to 4 K as depicted in Fig. 4(c). However, the initial negative offset of the \textit{$\rho^T_{yz}$} curve might be due to overestimation of the background Hall contribution during the fitting procedure. Importantly, the Hall signal exhibits linearity above $T_N$, thereby excluding multiband OHE and indicating that the residual Hall contribution below $T_N$ is linked to the competing magnetic interactions. 
\newline
In anisotropic systems such as EuAuBi, the Hall response is inherently sensitive to the crystallographic orientation and measurement geometry, as it originates from the coupling between the electronic structure and field-induced magnetic configurations. Furthermore, the THE in \textit{$\rho_{yz}$} configuration, but not in the \textit{$\rho_{zx}$} configuration, aligns with the magnetization behavior of EuAuBi. It is noteworthy that field-induced first-order metamagnetic transitions are observed only in the \textit{B}$\perp$\textit{z} configuration \cite{Takahashi2023SuperconductivityEuAuBi,Lipika2026}. Consequently, in electrical transport experiments, it is crucial to apply the magnetic field and measure the Hall voltage, both, in the \textit{B}$\perp$\textit{z} configuration to detect any THE. 
Notably, the extracted residual Hall contribution does not manifest as a broad feature but instead displays two distinct anomalies or peaks that coincide with the metamagnetic transitions observed in  magnetization. This suggests an abrupt reconstruction of magnetic texture under the applied field and temperatures rather than the stabilization of a single magnetic texture across the entire field and temperature range. Similar THE signal has been reported for EuCd$_2$, where multiple peaks corresponding to THE were observed, aligning with the metamagnetic transitions. These peaks were attributed to non-collinear spin textures localized at magnetic domain boundaries or phase coexistence regions \cite{Nishihaya2024TopologicalEuCd2}. In addition, THE signal linked to field-induced magnetic transitions have been reported in several Eu-based topological materials like EuAgAs, EuAuSb, EuCuAs, EuCuSb, etc, where they are attributed to the non-trivial or non-collinear spin configurations \cite{Ram2024, Wang2025, Roychowdhury2023InterplayEuCuAs, Malick2022}. Importantly, the observation
of a finite THE, closely correlated with metamagnetic transitions in EuAuBi, provides transport-based evidence supporting the presence of non-trivial real space magnetic textures in the system, consistent with the previous ac susceptibility studies \cite{Lipika2026}.
\newline
To gain a deeper understanding of the thermodynamic signatures associated with the field-induced magnetic textures, we performed specific heat measurements under a constant magnetic field. Fig. 5(a) shows the zero-field temperature-dependent specific heat \textit{C$_p$}(\textit{T}) of Pb-flux grown EuAuBi. A distinct \textit{$\lambda$}-like peak is observed around 4.8 K, indicative of a second-order antiferromagnetic-paramagnetic phase transition (as depicted in Fig. 5(a) inset). This is consistent with the \textit{M}(\textit{T}) measurements, shown in Fig. 3(a).  The experimentally obtained \textit{C$_p$} value at 200 K is 71.8 Jmol$^{-1}$K$^{-1}$, which closely aligns with the theoretical value obtained using the Dulong-Petit law, \textit{C$_p$} = 3\textit{N}R, where \textit{N} is the number of atoms per formula unit, (\textit{N} = 3 for EuAuBi) and R is the universal gas constant. 
\newline
To analyze the paramagnetic contribution, the experimental \textit{C$_p$}(\textit{T}) data above 30 K were fitted by considering both electronic and lattice contribution using the expression :
\begin{equation}
C(T) = \gamma T + a\, C_{\mathrm{Debye}}(T) + (1-a)\, C_{\mathrm{Einstein}}(T)
\end{equation}
where $\gamma$ is the Sommerfeld coefficient, a is the weighting factor. The first term represents the electronic contribution, while the second and third terms correspond to Debye and Einstein phonon contributions, respectively, given by :
\begin{equation}
C_{\mathrm{Debye}}(T) = 9 N R \left(\frac{T}{\Theta_D}\right)^3
\int_0^{\Theta_D/T}
\frac{x^4 e^x}{(e^x - 1)^2} \, dx
\end{equation}
\begin{equation}
C_{\mathrm{Einstein}}(T) = 3 N R
\left(\frac{\Theta_E}{T}\right)^2
\frac{e^{\Theta_E/T}}{\left(e^{\Theta_E/T}-1\right)^2}
\end{equation}
The fitting yields \textit{$\Theta$$_D$} = 136.58 K, \textit{$\Theta$$_E$} = 116.82 K, \textit{$\gamma$} = 0.1 mJ mol$^{-1}$ K$^{-2}$, a = 0.54. The very small electronic contribution might be due to semimetallic nature of EuAuBi, as also reported previously by DFT calculations \cite{Takahashi2023SuperconductivityEuAuBi}. The Debye temperature obtained using the specific heat analysis is in close agreement with that extracted using resistivity analysis in Fig. 3(c), indicating consistency between the thermodynamic and transport measurements. Using these parameters, we fitted the entire temperature range of \textit{C$_p$}(\textit{T}), as illustrated in Fig. 5(a). 
\newline
The magnetic contribution to specific heat, \textit{{C}$_{mag}$}, is obtained by subtracting the fitted phononic and electronic contribution using Eq. (1) from the experimental \textit{C$_p$}(\textit{T}) data. The magnetic entropy is then calculated as :
\begin{equation}
S_{\mathrm{mag}}(T) = \int_{0}^{T} \frac{C_{\mathrm{mag}}(T')}{T'} \, dT'
\end{equation}
Fig. 5(b) presents the temperature dependence of \textit{C$_{mag}$}/\textit{T} and \textit{S$_{mag}$}. The magnetic entropy exhibits a rapid increase up to \textit{T$_N$}, followed by a gradual rise until it saturates at higher temperature. The entropy saturates at approximately 15.5 J mol$^{-1}$ K$^{-1}$, which is close to the expected magnetic entropy \textit{R}ln(2\textit{S}+1) = \textit{R}ln8 J mol$^{-1}$ K$^{-1}$ for Eu$^{2+}$ (\textit{S} = 7/2). Notably, the magentic entropy is recovered around 25 K, while the \textit{T$_N$} $\sim$ 4.8 K, indicating the presence of short-range correlation above the long-range ordering temperature (\textit{T$_N$}). This behaviour is similarly observed in other Eu based compounds like EuMg$_2$Sb$_2$, EuCo$_2$P$_2$, EuMg$_2$Bi$_2$, etc \cite{Pakhira2022A-typeCrystals, Pakhira2020MagneticCrystals,Sangeetha2016EuCo2Antiferromagnet}.
\newline
\begin{figure}[t]
{
    \includegraphics[width=9cm]{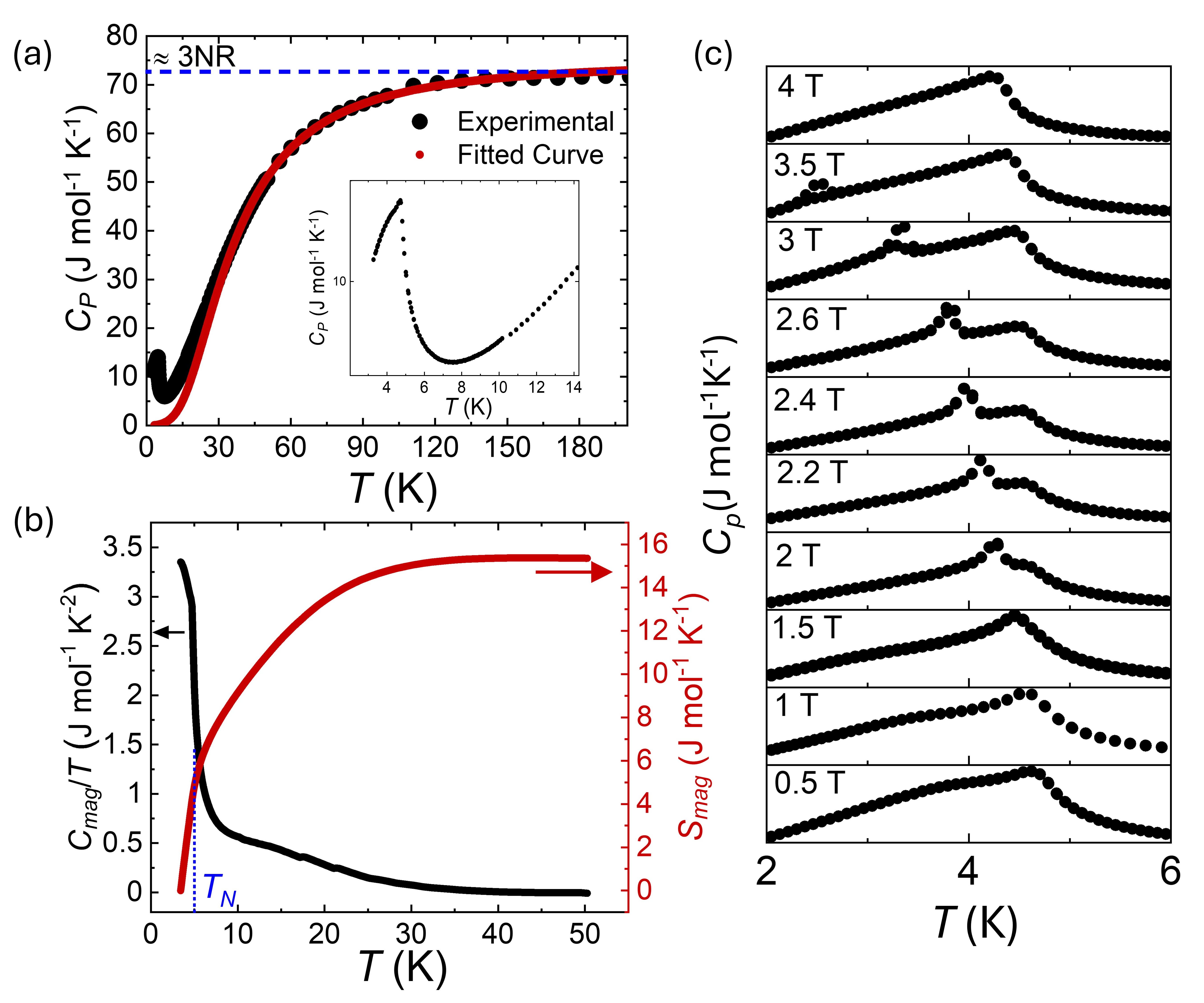}
    \caption{(a) Black curve is the experimental data of zero field temperature-dependent specific heat \textit{C$_p$} (\textit{T}) and red curve represents the fitting of experimental curve using contributions from electronic and phononic (Debye and Einstein) specific heats. (b)  Temperature-dependent plot of magnetic part of specific heat (C$_{mag}$/T) and entropy (S$_{mag}$. (c) Temperature-dependent specific heat at different magnetic fields. }
    \label{fig:my_label5}
    }
\end{figure}
The temperature dependence of \textit{C$_p$} measured under different applied magnetic fields is depicted in Fig. 5(c). Similar to the zero-field \textit{C$_p$}(\textit{T}), the $\lambda$ peak associated with the magnetic transition is observed around 4.8 K, corresponding to the antiferromagnetic transition. At 2 T, an additional peak appears alongside the $\lambda$ peak, which shifts towards a lower temperature as the field increases. Importantly, this peak arises in the same temperature and field window where the metamagnetic transitions and topological Hall contributions are observed. This finding suggests that the field-induced transition to non-trivial spin texture in EuAuBi single crystals is likely of first order in nature \cite{Lipika2026}. 
\newline
In conclusion, we have demonstrated that the Pb-flux growth is an effective method for achieving high-quality, impurity-free single crystals of magnetic topological semimetal EuAuBi, facilitating the investigation of its magnetic and transport properties. Unlike the Bi-flux grown samples, the Pb-grown crystals exhibit well-defined \textit{c}-axis orientation, and those grown under optimized conditions show no evidence of superconductivity. This finding indicates that the previously reported \cite{Takahashi2023SuperconductivityEuAuBi} superconducting transition originates from the extrinsic Au$_2$Bi impurity phases and is not intrinsic to EuAuBi. The \textit{c}-oriented crystals enable Hall measurements in an unexplored geometry, where we observe a finite residual Hall contribution emerging below the antiferromagnetic ordering temperature. This contribution appears within a narrow field–temperature window, coinciding with metamagnetic transitions in magnetization, anomalies in magnetoresistance, and additional features in field-dependent specific heat, suggesting a strong coupling between charge transport and field-induced magnetic reconstructions. The correlation between these transport and thermodynamic signatures provides compelling evidence for the emergence of non-collinear spin configurations, which give rise to a real-space Berry curvature contribution to the Hall response. Our results clarify the intrinsic ground state of EuAuBi and underscore the crucial role of the crystal orientation in revealing topological transport signatures in magnetic semimetals. Thus, our investigation establishes EuAuBi as a promising material platform for studying the interplay between magnetism, real-space topology, and electronic band topology.
\newline
\newline
We acknowledge Max Planck Society for the funding support under Max Plank-India partner group project, Science and Engineering Research Board, DST, Government of India, via grant no: CRG/2022/001826 and Defence Research and Development Organisation (DRDO), Ministry of Defence, Government of India (Project no: DFTM/033203/P/41/JATC-P2QP-17). Lipika would like to thank Prime Minister Research fellowship grant (PMRF Id: 1402122) for the research support and fellowship. Authors also thank the Central Research Facility (CRF), IIT Delhi, and the Nanoscale Research Facility at IIT Delhi for providing the materials characterization facility, like PPMS, MPMS, EDX, XRD, etc.
\bibliographystyle{apsrev4-2}
\bibliography{bibliography.bib}
\end{document}